\documentclass[12pt]{article}
\usepackage{graphicx}
\makeatletter
\@addtoreset{equation}{section}

\makeatother

\begin{document}

\begin{titlepage}
\null
\begin{flushright}
UT-02-54
\\
hep-th/0211148
\\
November, 2002
\end{flushright}

\vskip 1.5cm
\begin{center}

  {\LARGE Towards Noncommutative Integrable Systems}

\lineskip .75em
\vskip 2cm
\normalsize

  {\large Masashi Hamanaka\footnote{On leave of absence
from university of Tokyo, Hongo.
New e-mail: hamanaka@hep1.c.u-tokyo.ac.jp} and 
Kouichi Toda\footnote{e-mail: kouichi@yukawa.kyoto-u.ac.jp} }

\vskip 2cm

  {\it Institute of Physics, University of Tokyo, Komaba,\\
              Meguro-ku, Tokyo 153-8902, Japan}

\vskip 0.5cm

  {\it Department of Mathematical Physics,
Toyama Prefectural University,\\
Toyama, 939-0398, Japan}

\vskip 1.5cm

{\bf Abstract}

\end{center}

We present a powerful method to generate various equations
which possess the Lax representations
on noncommutative $(1+1)$ and $(1+2)$-dimensional spaces. 
The generated equations contain noncommutative integrable
equations obtained by using the bicomplex method
and by reductions of the noncommutative
(anti-)self-dual Yang-Mills equation.
This suggests that
the noncommutative Lax equations would be integrable
and be derived from reductions of the noncommutative 
(anti-)self-dual Yang-Mills equation,
which implies the noncommutative version of Richard Ward conjecture.
The integrability and the relation to 
string theories are also discussed.

\end{titlepage}

\clearpage

\baselineskip 6mm

\section{Introduction}

Non-Commutative (NC) gauge theories have been studied intensively 
for the last several years 
and succeeded in revealing various aspects of gauge theories
in the presence of background magnetic fields \cite{DoNe}. 
Especially,
NC solitons play crucial roles in the study of D-brane
dynamics, such as tachyon condensation \cite{Harvey}.
 
NC spaces are characterized by the noncommutativity of
the coordinates:
\begin{eqnarray}
\label{nc_coord}
[x^i,x^j]=i\theta^{ij},
\end{eqnarray}
where $\theta^{ij}$ are real constants.  
This relation looks like the canonical commutation
relation in quantum mechanics
and leads to ``space-space uncertainty relation.''
Hence
the singularity which exists on commutative spaces could resolve
on NC spaces.
This is one of the distinguished features of NC theories
and gives rise to various new physical objects.
For example, even when the gauge group is $U(1)$,
instanton solutions still exist \cite{NeSc}
because of the resolution of the small instanton singularities 
of the complete instanton moduli space \cite{Nakajima}.

NC gauge theories are naively realized from
ordinary commutative theories just by replacing
all products of the fields with star-products.
In this context, NC theories are
considered to be deformed theories from commutative ones
and look very close to the commutative ones.

Under the deformation,
the anti-self-dual
\footnote{There is essentially no difference 
between self-dual case and anti-self-dual case,
which allows us to restrict ourselves to anti-self-dual case.}
(ASD) Yang-Mills equations
could be considered to preserve the integrability
in the same sense as in commutative cases \cite{KKO, Nekrasov}.
On the other hand, with regard to typical integrable equations
such as the Kadomtsev-Petviasfvili (KP) equation \cite{KaPe},
the naive NC extension generally destroys the integrability.
There is known to be a method, the bicomplex method, to yield
NC integrable equations which have infinite number of conserved quantities
\cite{Bicomplex, DiMH, GrPe}.

On commutative spaces,
there is a fascinating relationship
between the ASD Yang-Mills equation and
many other lower-dimensional integrable equations,
that is, 
almost all integrable equations on $(1+1)$ and $(1+2)$-
dimensional spaces can be derived from 
the 4-dimensional ASD Yang-Mills equation by reductions.
This is first conjectured by R. Ward \cite{Ward}
(Ward conjecture) and can be said to be positively 
confirmed now \cite{Conj}.

The reduced equations always possess 
the Lax representations \cite{Lax}
because the original ASD Yang-Mills equation
also possesses the Lax representation in a wider sense.
The Lax representations are common in many integrable
equations. That is one of the reasons 
why Ward conjecture is reasonable and true. 
The equation which possesses the Lax representation
is called the Lax equation.

The NC ASD Yang-Mills equation 
also possesses the Lax representation as the NC-deformed form.
The reduced equations can also be NC Lax equations.
Since the NC ASD Yang-Mills equation is integrable,
it is expected that the reduced Lax equations
are also integrable as in commutative case.
This implies the NC version of Ward conjecture
and would be expect to open the door to a new
study area of integrable systems.
However it is very hard to find NC Lax equations
from reductions of the NC ASD Yang-Mills equation and so on.
The successful examples seem to be found only in \cite{Legare}.

\begin{figure}[htbn]
\begin{center}
\includegraphics[width=11cm]{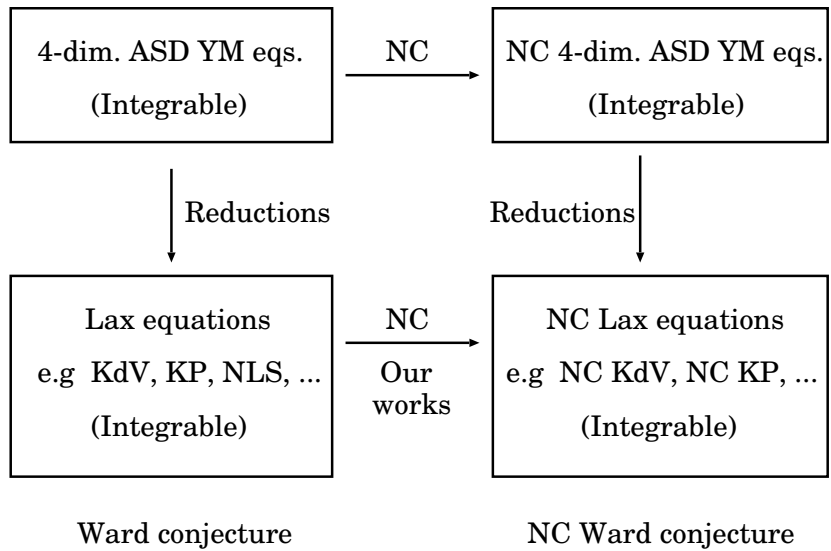}
\caption{NC Ward Conjecture}
\label{ward}
\end{center}
\end{figure}

In this letter, we discuss
NC extensions of wider class of integrable equations
which are expected to preserve the integrability.
First, we present a powerful method 
to generate various NC Lax equations.
Then we discuss the relationship between
the generated Lax equations and the NC integrable equations 
obtained from the bicomplex method
and from reductions of the NC ASD Yang-Mills equation.
All the results are consistent and 
some aspects of the integrability are proved, such as
the linearizability of the NC Burgers equation,
the existence of NC versions of KdV, KP and Burgers 
hierarchies and so on.
These results strongly suggest that
the NC Lax equations would be unique and integrable. 
Hence it is natural to
propose the following conjecture which contains 
the NC version of Ward conjecture:
{\it many NC Lax equations would be integrable
and be obtained from reductions of the NC ASD Yang-Mills equation}.
(See Fig. \ref{ward}.)
In the final section, more on the integrability and
the relation to string theories are discussed.

\vspace{3mm}

\section{Noncommutative Lax Equations}

\subsection{The Lax-Pair Generating Technique}

In commutative cases, 
Lax representations are
common in many known integrable equations
and fit well to the discussion of reductions 
of the ASD Yang-Mills equation.
Here we look for the Lax representations on NC spaces.
First we introduce how to find Lax representations 
on commutative spaces.

An integrable equation which possesses the Lax representation
can be rewritten as the following equation:
\begin{eqnarray}
[L,T+\partial_t]=0,
\label{lax}
\end{eqnarray}
where $\partial_t:=\partial/\partial t$.
This equation and the pair of operators $(L,T)$ are 
called the {\it Lax equation} and the {\it Lax pair}, respectively.

The NC version of the Lax equation (\ref{lax}),
the {\it NC Lax equation}, is easily defined
just by replacing the products of fields in $L$ and $T$ with the star products.
The star product is defined for usual functions $f(x)$ and $g(x)$ by
\begin{eqnarray}
f\star g(x):=
\exp{\left(\frac{i}{2}\theta^{ij}\partial^{(x^\prime)}_i
\partial^{(x^{\prime\prime})}_j\right)}
f(x^\prime)g(x^{\prime\prime})\Big{\vert}_{x^{\prime}
=x^{\prime\prime}=x.}
\end{eqnarray}
The star product has associativity: 
$f\star (g\star h)=(f\star g)\star h$,
and reduces to the ordinary product 
in the commutative limit $\theta^{ij}\rightarrow 0$.
The modification of the product  makes the ordinary
coordinate ``noncommutative,'' 
which means : $[x^i,x^j]_\star :=x^i\star x^j-x^j\star x^i
=i\theta^{ij}$.
We note that the coordinates and the functions
themselves are the same as commutative ones and 
take c-number values. Hence the differentiations
and the integrations of them are also the same as
commutative ones.
The effects of the NC deformation
appear only in the products of the coordinates
in the functions.

In this letter, we look for the NC Lax equation
whose operator $L$ is a differential operator.
In order to make this study systematic,
we set up the following problem :

\vspace{3mm}
\noindent
{\bf Problem :}
For a given operator $L$, 
find a corresponding operator $T$
which satisfies the Lax equation (\ref{lax}).
\vspace{3mm}

This is in general very difficult to solve.
However if we put an ansatz on the operator $T$,
then we can get the answer 
for wide class of Lax pairs
including NC case.
The ansatz for the operator $T$ is of the following type:

\vspace{3mm}
\noindent
{\bf Ansatz for the operator $T$ :}
\begin{eqnarray}
\label{ansatz}
T=\partial_i^n L^m+T^\prime.
\end{eqnarray}
\noindent
Then the problem for $T$ reduces to that for $T^\prime$.
This ansatz is very simple, however, very strong
to determine the unknown operator $T^\prime$.
In this way, we can get the Lax pair $(L,T)$,
which is called, in this letter, the
{\it Lax-pair generating technique}.

In order to explain it more concretely,
let us consider the Korteweg-de-Vries (KdV) 
equation \cite{KdV} on commutative $(1+1)$-dimensional space
where the operator $L$ is given 
by $L_{\scriptsize\mbox{KdV}}:=\partial_x^2+u(t,x)$.

The ansatz for the operator $T$ is given by
\begin{eqnarray}
T=\partial_x L_{\scriptsize\mbox{KdV}} +T^\prime,
\label{ansatz_KdV}
\end{eqnarray}
which corresponds to $n=m=1$ and $\partial_i=\partial_x$ in
the general ansatz (\ref{ansatz}).
This factorization 
was first used to find wider class of
Lax pairs in higher dimensional case \cite{ToYu}.

The Lax equation (\ref{lax}) leads to the equation for 
the unknown operator $T^\prime$:
\begin{eqnarray}
[\partial_x^2+u,T^\prime]=u_x\partial_x^2+u_t+uu_x,
\label{Tpr}
\end{eqnarray}
where $u_x:=\partial u/\partial x$ and so on.
Here we would like to 
delete the term $u_x\partial_x^2$ in the RHS of (\ref{Tpr})
so that this equation
finally reduces to a differential equation.
Therefore the operator $T^\prime$ could be taken as
\begin{eqnarray}
T^\prime=A \partial_x+B,
\end{eqnarray}
where $A, B$ are polynomials of $u, u_x, u_t, u_{xx},$ etc.
Then the Lax equation becomes $f\partial_x^2+g\partial_x+h=0$.
{}From $f=0, g=0$, we get\footnote{
In the present letter, we set the integral constants zero.} 
\begin{eqnarray}
A=\frac{u}{2},~~~B=-\frac{1}{4}u_x,
\end{eqnarray}
that is,
\begin{eqnarray}
T=\partial_x^3+\frac{3}{4}u_x+\frac{3}{2}u\partial_x.
\end{eqnarray}
Finally $h=0$ yields the Lax equation, the KdV equation:
\begin{eqnarray}
u_t+\frac{3}{2}uu_x+\frac{1}{4}u_{xxx}=0.
\end{eqnarray}

In this way, we can generate a wide class of Lax equations 
including higher dimensional integrable equations \cite{ToYu}.
For example, $L_{\scriptsize\mbox{mKdV}}
:=\partial_x^2+v(t,x)\partial_x$ and 
$L_{\scriptsize\mbox{KP}}
:=\partial_x^2+u(t,x,y)+\partial_y$ give rise to 
the modified KdV equation and the KP equation, respectively 
by the same ansatz (\ref{ansatz_KdV}) for $T$.
If we take $L_{\scriptsize\mbox{BCS}}:=\partial_x^2+u(t,x,y)$ and 
the modified ansatz $T=\partial_y L_{\scriptsize\mbox{BCS}}
+T^\prime$,
then we get the Bogoyavlenskii-Calogero-Schiff (BCS) 
equation \cite{BoSc}.\footnote{The multi-soliton solution 
is found in \cite{YTSF}.}

Good news here is that this technique is also applicable to
NC cases.

\subsection{Some Results}

We present some results by using the Lax-pair generating technique.
First we focus on NC $(1+2)$-dimensional
Lax equations where the time and space coordinates 
are denoted by $t$ and $(x,y)$, respectively.
Let us suppose that the noncommutativity is basically
introduced in the space directions, that is, $[x,y]=i\theta$.
\begin{itemize}

\item The NC KP equation \cite{Paniak} :

The Lax operator is given by
\begin{eqnarray}
L_{\scriptsize\mbox{KP}}=\partial_x^2+u(t,x,y)+\partial_y
=:L_{\scriptsize\mbox{KP}}^\prime+\partial_y.
\end{eqnarray}
The ansatz for the operator $T$ is the same as commutative case:
\begin{eqnarray}
T=\partial_x L_{\scriptsize\mbox{KP}}^\prime+T^\prime.
\end{eqnarray}
Then we find
\begin{eqnarray}
T^\prime=\frac{1}{2}u\partial_x-\frac{1}{4}u_x
-\frac{3}{4}\partial_x^{-1}u_y,
\end{eqnarray}
and the NC KP equation:
\begin{eqnarray}
\label{ncKP}
u_t+\frac{1}{4}u_{xxx}+\frac{3}{4}(u_x\star u+u\star u_x)
+\frac{3}{4}\partial_x^{-1}u_{yy}
+\frac{3}{4}[u,\partial_x^{-1} u_y]_\star
=0,
\end{eqnarray}
where $\partial_x^{-1}f(x):=\int^x dx^\prime f(x^\prime),
~u_{xxx}=\partial^3 u/\partial x^3$ and so on.
This coincides with that in \cite{Paniak}.
There is seen to be a nontrivial 
deformed term $[u,\partial_x^{-1} u_y]_\star$
in the equation (\ref{ncKP}) which vanishes 
in the commutative limit.
In \cite{Paniak}, the multi-soliton solution is found
by the first order to small $\theta$ expansion,
which suggests that this equation would be considered as an 
integrable equation.

\item The NC BCS equation :

This is obtained by following 
the same steps as in the commutative case.
The new equation is 
\begin{eqnarray}
&&u_t+\frac{1}{4}u_{xxy}+\frac{1}{2}(u_y\star u+u\star u_y)
+\frac{1}{4}u_x\star (\partial_x^{-1} u_y)\nonumber\\
&&+\frac{1}{4}(\partial_x^{-1}u_{y})\star u_x
+\frac{1}{4}[u,\partial_x^{-1}[u,\partial_x^{-1}u_{y}]_\star]_\star
=0,
\end{eqnarray}
whose Lax pair and the ansatz are
\begin{eqnarray}
L_{\scriptsize\mbox{BCS}}&=&\partial_x^2+u(t,x,y),\nonumber\\
T&=&\partial_y L_{\scriptsize\mbox{BCS}}+T^\prime,\nonumber\\
T^\prime&=&\frac{1}{2}(\partial_x^{-1}u_y)\partial_x
-\frac{1}{4}u_y
-\frac{1}{4}\partial_x^{-1}[u,\partial_x^{-1}u_y]_\star.
\end{eqnarray}
This time, a non-trivial term is found even in the operator $T$.

\end{itemize}

We can generate many other NC Lax equations 
in the same way. (For explicit examples, see \cite{Toda}.)
Moreover if we introduce the noncommutativity 
into time coordinate as $[t,x]=i\theta$,
we can construct NC $(1+1)$-dimensional integrable equations.
The introduction of noncommutativity in the time direction
makes it hard or ambiguous to define the initial value 
problem and the integrability. 
We discuss this point for explicit examples as follows.

For example, the NC KdV equation is
\begin{eqnarray}
u_t+\frac{3}{4}(u_x\star u +u\star u_x)+\frac{1}{4}u_{xxx}=0,
\end{eqnarray}
which coincides with that derived 
by using the bicomplex method \cite{DiMH}
and by the reduction from NC KP equation (\ref{ncKP})
setting the fields $y$-independent: $\partial_y u =0$.
Here we reintroduce the noncommutativity 
as $[t,x]=i\theta$.\footnote{We note that this reduction is formal
and the noncommutativity here contains subtle points in
the derivation from the $(2+2)$-dimensional
NC ASD Yang-Mills equation by reduction
because the coordinates $(t,x,y)$
originate partially from the parameters 
in the gauge group of the NC Yang-Mills theory \cite{Conj}.
We are grateful to T.~Ivanova for pointing out this point to us.}

If we take the ansatz $T
=\partial_x^n L_{\scriptsize\mbox{KdV}}+T^\prime$,
the $(n+2)$-th NC KdV hierarchy equations are derived.
(The 3-th NC KdV hierarchy equation is just the NC KdV equation.)
This time, we can find the explicit form of $T^\prime$
as $T^\prime=\sum_{l=0}^{n} A_l \partial_x^{n-l}$
where $A_l$ are homogeneous polynomials of $u,u_x,u_{xx}$ 
and so on, whose degrees are $[A_l]=l+2$.
The important point is that these unknown 
polynomials are determined one by one 
as $A_0=(n/2)u$ and so on, which guarantees the existence
of the NC KdV hierarchy.
Moreover it is interesting that the $2n$-th 
NC KdV hierarchy equations seem to be also trivial
as commutative case.
These results strongly suggest the possibility
of NC extension of Sato theory \cite{Sato},
which will be discussed more in detailed \cite{work}.
The explicit and detailed discussions of the NC KdV hierarchy
are found in \cite{Toda}.

As one of the new Lax equations, 
the NC Burgers equation is obtained:
\begin{eqnarray}
u_t-\alpha u_{xx}+(1-\alpha-\beta)u\star u_x
+(1+\alpha-\beta)u_x\star u=0.
\label{nc_burgers}
\end{eqnarray}
We can linearize it by the following two kind of
NC Cole-Hopf transformations 
\cite{HaTo, MaPa}:\footnote{After submitting
the present letter and our paper \cite{HaTo}, 
we were aware of the paper
\cite{MaPa} by L. Martina and O. K. Pashaev
on arXiv e-print server,
which contains some overlaps with ours.}
\begin{eqnarray}
\label{ch1}
u=\psi^{-1}\star \psi_x,
\end{eqnarray}
only when $1+\alpha-\beta=0$, and
\begin{eqnarray}
\label{ch2}
u=-\psi_x\star \psi^{-1},
\end{eqnarray}
only when $1-\alpha-\beta=0$.
The linearized equation is the NC
diffusion equation
\begin{eqnarray}
\psi_t=\alpha\psi_{xx},
\end{eqnarray}
which is solvable via the Fourier transformation.
Hence the NC Burgers equation 
is really integrable.

It is very interesting that
an equation on NC $(1+1)$-dimensional spaces
with infinite number of time derivatives
can be mapped to the NC diffusion equation
with the first derivative of time.
This shows that the NC Burgers equation,
an infinite differential equation,
is actually integrable!
In the exact solutions, 
non-trivial effects of the NC-deformation
are also seen \cite{HaTo, MaPa}.

The transformations (\ref{ch1}) and (\ref{ch2})
are analogy of the commutative
Cole-Hopf transformation $u=\partial_x \log \psi$ \cite{CoHo}.
This success makes us expect the possibility of
NC extensions of the Hirota's method \cite{Hirota}
which is a strong and direct way to construct
the exact multi-soliton solutions.

The ($n$-th) NC Burgers hierarchy equations are also obtained
by taking the ansatz $T=\partial_x^n 
L_{\scriptsize\mbox{Burgers}}+T^\prime$ \cite{HaTo}.
The NC KP and BCS hierarchies could be also derived 
by similar ansatz.
Hence the reductions of $(1+2)$-dimensional
NC equations would give rise to $(1+1)$-dimensional
NC equations, which is a part of 
the NC Sato theory \cite{work}.

Moreover, all of the NC integrable equations derived 
from the bicomplex method
can also be obtained by our method.
The bicomplex method guarantees the existence of
infinite number of conserved topological quantities.

All these results suggest that NC Lax equations would possess 
the integrability. 

\vspace{3mm}

Let us here comment on the multi-soliton solutions.
First we note that 
if the field is holomorphic, that is, $f=f(x-vt)=f(z)$,
then the star product reduces to the ordinary product:
\begin{eqnarray}
f(x-vt)\star g(x-vt)=f(x-vt)g(x-vt).
\end{eqnarray}
Hence the commutative
multi-soliton solutions where all the solitons
move at the same velocity always satisfy
the NC version of the equations.
Of course, this does not mean that
the equations possess the integrability.
The integrability is characterized in this context by
the existence of multi-soliton solution with 
non-trivial scattering processes.

The comprehensive list and the more detailed discussion 
are reported later soon.

\section{Comments on the Noncommutative Ward Conjecture}

In commutative case, it is well known that 
many integrable equations could be
derived from symmetry reductions of the four-dimensional
ASD Yang-Mills equation \cite{Conj},
which is first conjectured by R.Ward \cite{Ward} 
({\it Ward conjecture}).

Even in NC case,
the corresponding discussions are possible and interesting.
The NC ASD Yang-Mills equation also has Yang's
form \cite{Nekrasov, Nekrasov2} and many other similar
properties to commutative cases \cite{KKO}.
A simple reduction to three dimension
yields the NC Bogomol'nyi equation
which has the exact monopole solutions and can be rewritten 
as the non-Abelian Toda lattice equation \cite{Nekrasov, GrNe}.
It is interesting that a discrete structure appears.

Moreover M.Legar\'e \cite{Legare} succeeded in some reductions
of the $(2+2)$-dimensional NC ASD Yang-Mills equation,
which coincide with our results 
and those by using the bicomplex method.
The NC Burgers equation is also derived
from the ASD equation in NC $U(1)$ Yang-Mills theory,
which is discussed in the revised version of \cite{HaTo}.
In NC theories, the $U(1)$ part is important and
the parameters in NC Burgers equation (\ref{nc_burgers}) actually
come from the commutator part such as $[A_\mu,A_\nu]$
in the field strength of the original NC Yang-Mills theory,
which becomes trivial in the commutative limit.
These results strongly suggest that the NC deformation
is unique and integrable
and the Ward conjecture still holds on NC spaces.

In four-dimensional Yang-Mills theory,
the NC deformation resolves the small instanton 
singularity of the (complete) instanton moduli space
and gives rise to a new physical object, the $U(1)$ instanton.
Hence the NC Ward conjecture would imply that
the NC deformations of lower-dimensional integrable equations
might contain new physical objects
because of the deformations of the solution spaces in some case.

\section{Conclusion and Discussion}

In the present letter, 
we found a powerful method to
find NC Lax equations which
is expected to be integrable.
The simple, but mysterious ansatz (\ref{ansatz}) 
plays an important role
and actually gives rise to various new NC Lax equations.
Finally we pointed out that some reductions of the NC
ASD Yang-Mills equation gives rise to
NC integrable equations including our results.

Now there are mainly three methods to yield NC integrable
equations:
\begin{itemize}
\item Lax-pair generating technique
\item Bicomplex method
\item Reduction of the ASD Yang-Mills equation
\end{itemize}
The interesting point is that 
all the results are consistent
at least with the known NC Lax equations, 
which suggests the existence and 
the uniqueness of the NC deformations of integrable equations
which preserve the integrability.

Though we can get many new NC Lax equations,
there need to be more discussions so that
such study should be fruitful as integrable systems.
First, we have to clarify whether the NC Lax equations
are really good equations in the sense of integrability, that is,
the existence of many conserved quantities or
of multi-soliton solutions, and so on.
All of the previous studies including our works
strongly suggest that this should be true.
Second, we have to reveal the physical meaning of such equations.
If such integrable theories can be embedded in string
theories, there would be fruitful interactions
between the both theories,
just as between the (NC) ASD Yang-Mills equation and
D0-D4 brane system (in the background of NS-NS $B$ field).

For NC $(1+1)$-dimensional equations,
the initial value problem is even hard to define
and the meaning of the integrability may be vague.
However we have got such kind of {\it integrable} equations,
for example, the NC Burgers equation which is {\it linearizable}.
This is somewhat surprising and it is also
interesting to reveal how the integrability is
defined or how the initial value problem is solved
with infinite number of time derivatives.
The mechanism may relate the physical meaning
of the introduction of the noncommutativity.

We believe that the systematic and co-supplement studies 
of them will pioneer a new area of integrable systems and 
perhaps string theories.

\vskip5mm\noindent
{\bf Note added}
\vskip2mm

\noindent
We were informed by O. Lechtenfeld 
that the $(2+2)$-dimensional NC ASD Yang-Mills equation and
some reductions of it can be embedded \cite{LPS, LePo}
in $N=2$ string theory \cite{OoVa},
which guarantees that such directions would have a physical
meaning and might be helpful to understand new aspects of 
the corresponding string theory.

\vskip7mm\noindent
{\bf Acknowledgments}
\vskip2mm

\noindent
We would like to thank the YITP at Kyoto University 
for the hospitality during the YITP
workshop YITP-W-02-04 on ``QFT2002''
and our stay as atom-type visitors.
M.H. is also grateful to M.~Asano, M.~Kato and I.~Kishimoto 
for useful comments.
The authors thank T.~Tsuchida and anonymous referees 
for careful reading of this
manuscript and valuable remarks.
The work of M.H. was supported in part 
by the Japan Securities Scholarship Foundation (\#12-3-0403).

\end{document}